# Perspective: Probing elasto-quantum materials with X-ray techniques and in situ anisotropic strain

Han Zhang[1†], Joshua J. Sanchez[2†], Jiun-Haw Chu[3*] and Jian Liu[4*]

[1] Changzhou University, Changzhou, Jiangsu, China, 213001
[2] Massachusetts Institute of Technology, Cambridge, MA; USA, 02139
[3] Department of Physics, University of Washington, Seattle, WA, USA; 98195
[4] Department of Physics and Astronomy, University of Tennessee; Knoxville, TN, USA, 37996

[†] Equal contribution
[*] Corresponding authors
E-mail: jianliu@utk.edu (J.L.), jhchu@uw.edu (J.-H. C.)

## Abstract

Anisotropic lattice deformation plays an important role in the quantum mechanics of solid state physics. The possibility of mediating the competition and cooperation among different order parameters by applying in situ strain/stress on quantum materials has led to discoveries of a variety of elasto-quantum effects on emergent phenomena. It has become increasingly critical to have the capability of combining the in situ strain tuning with X-ray techniques, especially those based on synchrotrons, to probe the microscopic elasto-responses of the lattice, spin, charge, and orbital degrees of freedom. Herein, we briefly review the recent studies that embarked on utilizing elasto-X-ray characterizations on representative material systems and demonstrated the emerging opportunities enabled by this method. With that, we further discuss the promising prospect in this rising area of quantum materials research and the bright future of elasto-X-ray techniques.

## 1. Introduction

The past decade has witnessed the development of what we refer to as "elasto-quantum mechanics", where the quantum physics in an increasing number of crystalline solid-state systems is effectively tuned and controlled by in situ mechanical deformation of the lattice. A milestone achieved by Chu et al. is the utilization of an anisotropic strain of <0.1% for resolving the relation between electron nematicity and structural phase transition in an iron arsenide superconductor [1]; as another milestone, Hick et al. reported the doubling of the superconducting transition temperature in $Sr_2RuO_4$ by introducing only ~0.2% anisotropic strain [2]. Over the years, several pioneering studies have demonstrated that the application of elastic anisotropic strain is remarkably powerful in realizing unparallel functional controls of various order parameters, stabilizing emergent phases of matter, and revealing new physics of quantum materials. This new technique has considerably transformed the fields of electronic nematicity [3], higher-order interactions [4], and unconventional superconductivity [2, 5, 6]. Notably, the applied anisotropic strain is often substantially smaller than 1%, but the impact is already energetically comparable to the application of a magnetic field at the order of 0.1 [7], 1 [8, 9], or 10 [10, 11] T in systems. With the discovery of more and more elasto-quantum materials, it has become increasingly important to resolve the actual microscopic lattice distortion and distinguish it from the macroscopic deformation because the former is directly linked to the underlying mechanism of the elasto-responses. In addition, there are rising demands for probing elasto-properties that are spatially modulated and temporally dependent, such as antiferromagnetism [7], charge order [12], spin dynamics [13], and orbital polarization [14].

Figure 1 compares different techniques that introduce structural distortions on a crystal lattice. Historically, strain engineering of quantum materials refers to the use of epitaxial growth of a material on structurally compatible but lattice-mismatched substrates. A major advantage of this approach is that the strain can go up to a few percent and the sample remains highly stable. Indeed, such techniques have enabled many emergent phenomena that do not occur in bulk crystals [15]. However, the size of the epitaxial strain is usually fixed, and the materials necessarily crystallize under the strain, raising the question of whether the film samples are comparable with the bulk crystals of the same materials. On the other hand, stress may be applied to bulk crystals by high-pressure techniques, as done in the case of the diamond anvil cell (DAC) [16] and the cubic anvil cell (CAC) [17]—this is a well-established field, where the structural, vibrational, electronic, and magnetic properties of materials are tuned over a wide range of pressure–temperature conditions [18]. A disadvantage of this approach is, however, that the stress, while controllable, cannot be continuously tuned in an in situ fashion and only compressive stress can be applied. The technical challenge of high-pressure studies with DAC or CAC is also related to the fact that the sample environment is extremely delicate with limited space and can only be combined with a few measurement techniques. X-ray scattering/spectroscopy is one of the compatible techniques and has been widely employed in high-pressure studies because of its scattering geometry and the capability of measuring small samples. The same strategy can be extended to elastic strain/stress studies by replacing the DAC/CAC with an in situ strain cell, which should even simplify the setup and make the sample environment more flexible thanks to the open design of the strain cells. Moreover, the applied strain/stress can be compressive or tensile by electrically switching the voltage on the piezo actuator of the strain cell. Such technique could be of particular interest in the emerging field of low-dimensional materials where symmetry-breaking phase transitions often play an important role in determining their unique properties [19, 20].

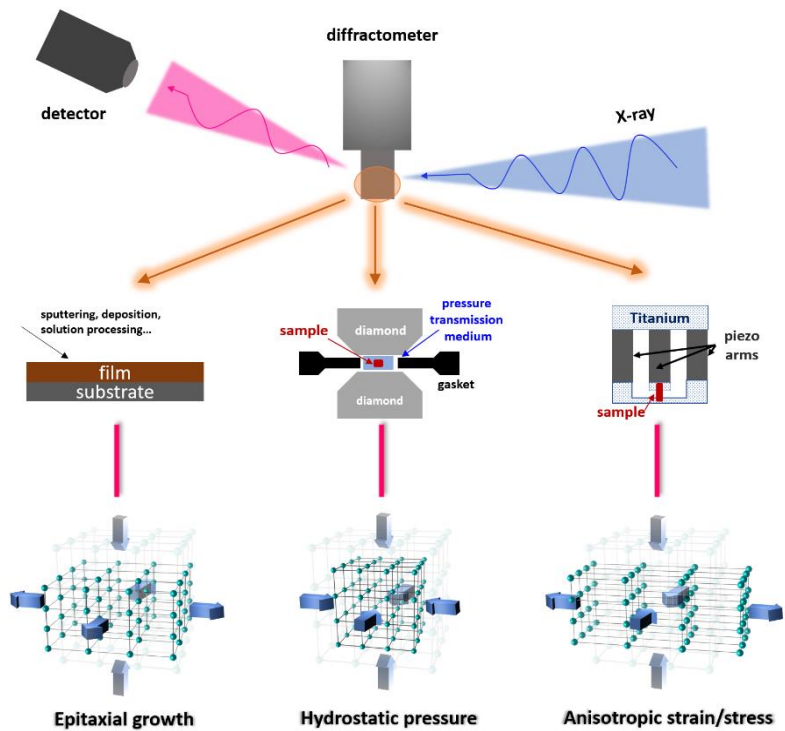

**Figure 1.** Schematic of different techniques that introduce structural distortion with in situ X-ray measurement. The effects on the crystal lattice are illustrated at the bottom.

The piezoelectricity-based strain technique was developed by Chu et al. over an iron based superconductor[1]. In this setup, a single crystal sample was glued to a piezo actuator. When an external voltage was applied, the piezo actuator expanded or contracted, in turn generating a uniaxial strain on the sample. The advantage of this approach is that the anisotropic strain can be in situ tuned under cryogenic conditions by adjusting the voltage of the piezo actuator. Thus, measurements can be made under a continuous scan of the applied strain in analogy to the magnetic field scan commonly used for measuring magnetic properties. This is in contrast to conventional techniques, such as mechanical clamping of the sample. An improved setup was developed by Hicks et al. by implementing a strain cell configuration comprising three piezo arms. This design is now widely used as the parallel geometry to achieve larger strains and cancel out the thermal expansion of the piezo material [21] (right panel of figure 1).

The applied elastic strain is usually monitored by measuring the sample deformation via a strain gage, which is usually a foil gage or a high-precision capacitor. This measurement is based on the presumption that the distance traveled by the piezo actuators is the displacement of the sample shape, which is however not necessarily the same as the actual change in the lattice structure. Strictly speaking, one only applies stress not strain because strain is the response to stress, though it is often convenient to use strain as a proxy for stress when the focus is on the elasto-responses of the electric and magnetic properties. Therefore, a direct probe for precise lattice distortion is urgently needed. Furthermore, in some specified quantum systems where the elasto-quantum effects are manifested as the detwinning of structural/magnetic domains [7, 11] or emergent charge/spin ordering [22, 23], a direct examination of the lattice, charge, and spin degree of freedom will greatly aid the understanding of the underlying mechanism. These are areas where X-ray techniques, such as X-ray diffraction, are known to shine, especially when performing the measurements at synchrotron facilities. The integration of in situ strain control with X-ray techniques is therefore highly desired.

Although X-ray offers an ideal solution to many of the scientific issues, incorporating the elastic anisotropic strain technique into an X-ray scattering/spectroscopy sample environment is a highly nontrivial technical task. Although the strategy is similar to that adopted for high-pressure experiments, the geometry of the strain cell holder and sample orientation relative to the diffractometer must be planned in accordance with the incoming beam direction/polarization/energy and the detector coverage to achieve the desired reflections within a wide range of the reciprocal space. All the cabling and wiring of the strain cell and the sample should be properly integrated in the cryocooler, especially when simultaneous measurements of the sample resistance are required, e.g., for comparing it with the elastoresistance. Careful technical considerations and developments are necessary to maximize the potential of elasto-X-ray techniques. In this perspective, we will introduce several examples of the recent efforts of the quantum materials community to probe emergent phenomena using elasto-X-ray techniques. We will discuss new opportunities in this emerging field of quantum materials research with focus on the future of elasto-X-ray characterization.

## 2. Discussion

### 2.1 Nematicity and superconductivity in iron-based superconductors: insights from X-ray diffraction

To date, high-temperature iron-based superconductors are the material system that has been most broadly characterized using combined strain and X-ray techniques. A rich variety of phases can be found across the diverse crystal structures and doping phase diagrams in this system, but a few features are common to most of these materials. An electronic nematic phase manifests as an electronically driven $C_4$ to $C_2$ rotational symmetry breaking. Thus breaking of the symmetry induces anisotropies in the crystal lattice, resistivity, spin fluctuation spectra, orbital occupations, etc. Suppression of this phase by chemical doping or pressure can often result in the emergence of superconductivity, leading to the conjecture that quantum critical nematic fluctuations may mediate or enhance unconventional Cooper pairing. Thus, nematicity has been intensely investigated.

The strain tunability of the nematicity and superconductivity in this system originates from the nematic coupling to the lattice. The crystal structure of all iron-based superconductors comprises square lattices of iron atoms with either arsenic (the pnictide materials) or selenium (the chalcogenide materials) atoms coordinated above and below the iron plane (figure 2(a)). For iron pnictides, a square plane of rare-earth or alkaline-earth elements (e.g., La, Ba, and Eu) separates the iron arsenide planes, whereas for iron chalcogenides, the FeSe planes are stacked without any intermediate layers. The $C_4$ rotational symmetry of the Fe square plane can be broken either by orthorhombic distortion in the Fe–As/Se direction ($B_{1g}$ symmetry) or the Fe–Fe direction ($B_{2g}$ symmetry). The nematic phase is associated with spontaneous orthorhombicity in the $B_{2g}$ symmetry channel, forming nematic/structural twin domains. These domains cause bulk probes to average over the two domains, obscuring spontaneous anisotropies within a single domain. By externally inducing $B_{2g}$ strain, the sample can be detwinned to a single domain state, enabling the characterization of the electronic anisotropies. However, further straining beyond the "perfectly detwinned" state will generally lead to further enhancement of nematicity and an increase in all electronic anisotropies, making it challenging to determine the spontaneous anisotropies and their temperature dependence.

The original motivation for combining strain tuning with X-ray diffraction in iron-based superconductors was to make the first precise measurement of spontaneous resistivity anisotropy within the nematic phase [11]. For this purpose, a sample of underdoped Ba(Fe$_{0.96}$Co$_{0.04}$)$_2$As$_2$ was prepared with transport wires and mounted over a gap between two titanium plates (figure 2(b)). After the material was cooled into the nematic phase, the presence of nematic twin domains was identified by two peaks in the X-ray diffraction (XRD) spectrum of the material. These peaks corresponded to the two lattice constants of the domains along the stress axis. The precise detwinning of the domains could then be monitored by tracking the intensity of these two peaks as strain was applied. Notably, the lattice constants of the domains were nearly constant across the detwinning strain range and only became strongly strain-tunable beyond full detwinning. This enables access to the barely detwinned monodomain state in the absence of additional lattice distortion.

The resistivity anisotropy was extracted from these barely detwinned points and tracked across temperature within the nematic phase. The ratio of spontaneous resistivity anisotropy to spontaneous orthorhombicity within the nematic phase was nearly constant. As both quantities are driven by microscopic nematicity, it was thus demonstrated that spontaneous resistivity anisotropy acts as a proxy for the nematic order parameter. In other words, this transport quantity behaves like a thermodynamic order parameter.

Another aspect of this work was the investigation of the structural responses to strain within the high-temperature tetragonal state. By tracking the change in the transverse and out-of-plane lattice constants as the strain-aligned lattice constant was tuned, the terms in the elastic modulus tensor were determined, enabling probing of the structural susceptibility beyond the zero-stress limit by an elasto-XRD technique. This technique was used to demonstrate that the shear modulus diverges toward zero as the material is cooled toward the nematic transition. A simultaneous measurement of the diverging elastoresistivity showed that the ratio of transport to structural anisotropy remained constant above the transition, with the same value as that in the nematic state. Thus, the transport–structural correspondence was maintained through the nematic transition.

A second study [24] using the same experimental design as the first aimed to leverage strain enhancement of nematicity to tune superconductivity. In this study, the sample was cooled to the superconducting state with zero resistance (R = 0) under zero strain (figure 2(c)). Under a large applied tension or compression, the nonzero resistivity reemerged, demonstrating the suppression of superconductivity by applying strain. Simultaneous XRD data showed that for small strains that only detwin the sample, the zero-resistance state is preserved. As the strain increases beyond the perfect detwinning point, the enhanced orthorhombicity and nematicity suppress the superconductivity. Thus, the static nematic order competes with the superconducting order. This ability to distinguish between the effects of detwinning and further lattice distortion is the key advantage of the elasto-XRD approach.

The competition between nematicity and superconductivity was further explored in a recent study on Co-doped $EuFe_2As_2$, a member of the 122 iron pnictide family [25]. In this study, stress was applied in the FeAs direction, inducing $B_{1g}$ orthorhombicity (figure 2(d)). This does not cause detwinning between the domains, but instead it indirectly tunes the magnitude of the spontaneous $B_{2g}$ orthorhombicity via higher-order couplings between lattice distortions of different symmetry [26]. Applying tension leads to a dramatic reduction in the spontaneous $B_{2g}$ orthorhombicity, indicating a suppression of nematicity, which is concomitant with a dramatic reduction in resistivity and an enhancement of superconductivity. These results demonstrate that such measurements can be used to precisely quantify the couplings between correlated phases (e.g., nematicity and superconductivity) with lattice distortions of different symmetries.

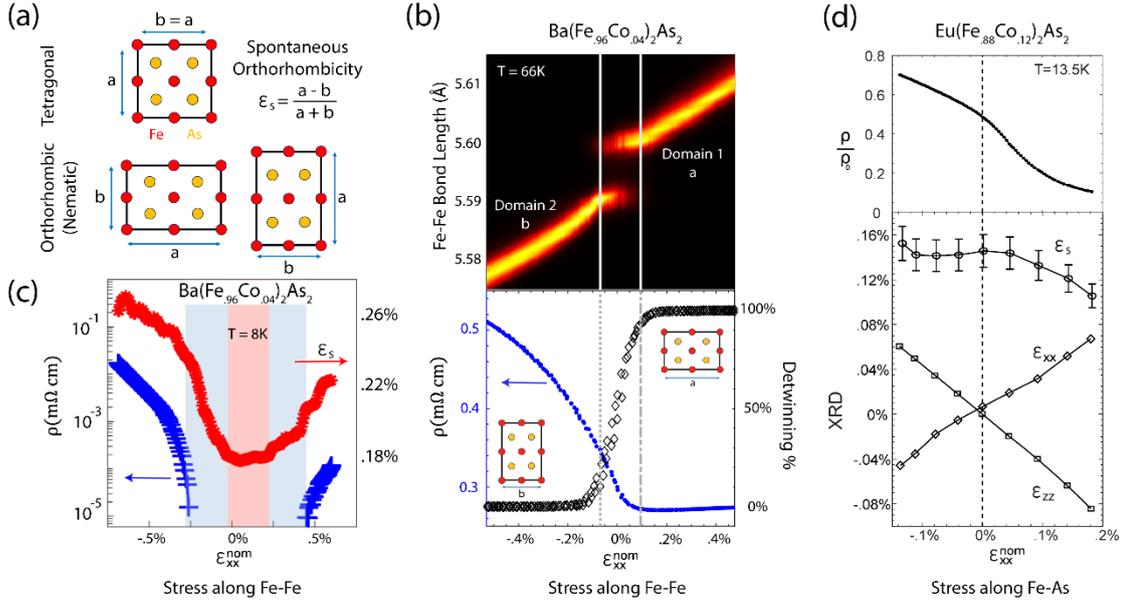

**Figure 2.** X-ray diffraction (XRD) spectra of nematicity in iron-based superconductors. **(a)** The FeAs plane of iron pnictide superconductors forms a square lattice in the high-temperature tetragonal phase. In the nematic phase, orthorhombic twin domains are formed via distortions in the Fe–Fe bonding direction. **(b)** For Ba(Fe$_{.96}$Co$_{.04}$)$_2$As$_2$ within the nematic phase, XRD spectrum in the Fe–Fe bonding direction under zero strain shows two intensity peaks, implying the presence of two orthorhombic domains. Small stress applied in this direction cause detwinning of the sample without increasing the lattice orthorhombicity, whereas larger stress further enhances the orthorhombicity of the monodomain state. **(c)** At T = 8 K, the same sample simultaneously exhibits nematic, antiferromagnetic, and superconducting properties, with ρ = 0 under zero strain. Simultaneous resistivity measurements and XRD in the Fe–Fe bonding direction reveal that small strains (red shading) do not increase the lattice orthorhombicity, whereas larger strains (blue shading) increase the orthorhombicity. (From [24]). **(d)** In Eu(Fe$_{.88}$Co$_{.12}$)$_2$As$_2$ at T = 13.5 K, the sample is nematic and is below the superconducting onset temperature, but it has not reached zero resistance. Applied tension suppresses nematic-driven orthorhombicity, as seen by the reduced value of $\varepsilon_S$. This enhances the superconductivity, yielding a reduced resistivity value (From [25]). The nominal strain $\varepsilon_{xx}^{nom}$ in (b–d) is the strain determined by the capacitance strain gage of the stress device.

## 2.2. Orbital polarization and magnetism probed using X-ray spectroscopy in iron-based superconductors

Combining strain and XRD with X-ray spectroscopy can provide abundant additional information about the coupling of orbital and magnetic quantities to the lattice. For instance, the above-discussed studies were based on structural orthorhombicity as a proxy for the nematic order parameter without characterizing its microscopic origin. Nematicity was deeply examined in a recent study [14] on FeSe. Here, the orbital occupation anisotropy between the Fe 3d$_{xz}$ and 3d$_{yz}$ orbitals was measured by the X-ray linear dichroism (XLD) technique at the Fe k-edge, which is the difference in absorption for incident X-rays linearly polarized in-plane parallel and transverse to the stress axis. Above the nematic transition point, orbital polarization clearly diverged toward the transition temperature, indicating an orbital driver for the transition. Within the nematic phase, the XLD result saturates at strains beyond full detwinning, indicating that orbital polarization does not simply result from lattice anisotropy. Thus, orbital polarization acts as a primary order parameter of the nematic phase. Accordingly, nematicity in FeSe, similar to the pnictides, is considered to be driven more strongly by an orbital instability than by spin fluctuations.

X-ray spectroscopy can also be used to probe magnetism via X-ray magnetic circular dichroism (XMCD), which is the difference between the absorption of left and right circular polarized X-rays. Unfortunately, Fe k-edge XMCD is relatively weak, and hard X-ray XMCD has not yet been used to characterize Fe magnetism in these materials. Nonetheless, XMCD is essential for solving outstanding problems related to the Eu-based iron pnictide EuFe$_2$As$_2$. In addition to Fe-based nematicity, antiferromagnetism, and superconductivity, the undoped and doped compounds exhibit diverse Eu-based magnetic orders, which can be probed via the strong Eu L-edge XMCD signal.

The first study [9] combining tunable strain and applied magnetic field with XMCD investigated the A-type antiferromagnetic (AFM) order in the parent compound EuFe$_2$As$_2$. This material has an unusually strong magnetostructural coupling owing to the interactions between the Eu and Fe moments. Hence, a magnetic field of only ~0.5

T can fully detwin the nematic domains. This field is two orders of magnitude less than that in other iron-based superconductors. However, this field detwinning poses a notable challenge for studying the field response of the Eu magnetic order. The strain was used to detwin the sample and maintain the single domain state fixed, while an applied magnetic field reoriented the Eu moments in different lattice directions. The energetic coupling terms between the Eu–Fe and Eu–Eu moments were quantitatively determined from the saturation field values in these directions. This study revealed that in addition to a large biquadratic coupling between the Eu and Fe moments, the Eu–Eu coupling itself was strongly anisotropic owing to the influence of the nematic conduction electrons. This is an unexpected side effect of nematicity. Notably, this study is the first to demonstrate the spin–flip transition of Eu moments, which had previously been obscured by field detwinning.

A second study [25] examined the optimal-doped compound $Eu(Fe_{0.88}Co_{0.12})_2As_2$, which has a superconducting transition onset temperature ($T_{SC}$) of 20 K and an Eu ferromagnetic transition temperature ($T_{FM}$) of 17 K. The coexistence of the two phases results in the rare phenomenon of ferromagnetic superconductivity. Rarer still, an in-plane applied magnetic field of ~0.2 T can induce a zero-resistance state at a temperature higher than the spontaneous zero-resistance temperature. By applying a field under different strain states and using XMCD to track the reorientation of the Eu moments from out-of-plane to in-plane orientations, it was shown that the peak field enhancement of superconductivity coincides with the complete in-plane saturation of the Eu moments. Thus, the combination of strain application and magnetic field with transport and XMCD measurements can help identify the mechanism of the field-induced superconductivity in this system.

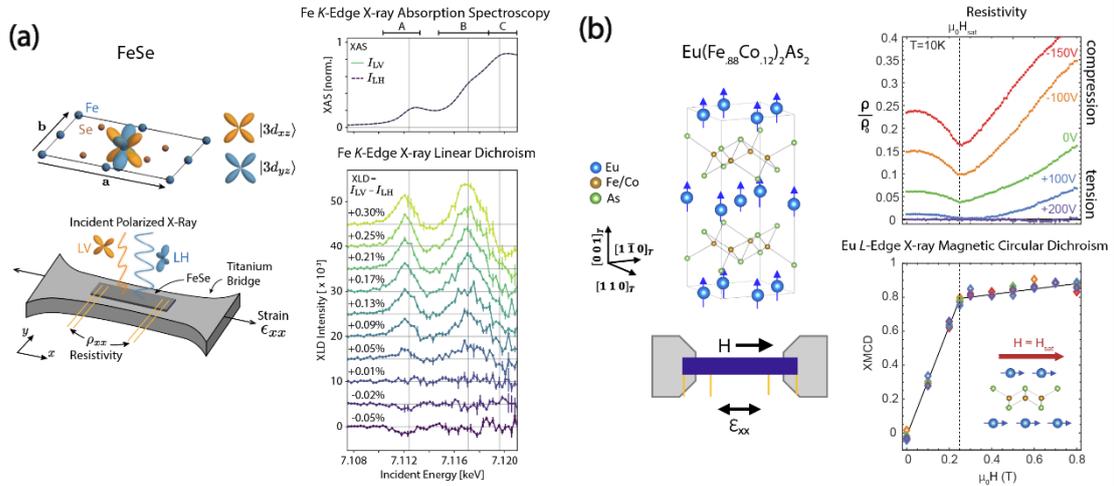

**Figure 3.** X-ray spectroscopy. **(a)** Unit cell of FeSe in the broken-symmetry nematic phase, with $3d_{xz}$ and $3d_{yz}$ orbitals for the centered Fe atom. Stress is applied in the Fe–Fe bonding direction using a titanium bridge that detwins the nematic domains. Linear vertical (LV)-polarized X-rays are more likely to create excitations from the $3d_{xz}$ orbitals, whereas linear horizontal (LH) X-rays cause excitations preferentially from the $3d_{yz}$ orbitals. While the raw X-ray absorption spectra (XAS) are virtually identical for LV and LH X-rays, the X-ray linear dichroism (XLD, taken as the difference of LV and LH results) exhibits strong strain tunability as the domains are detwinned, indicating large orbital polarization within the nematic phase (From [14]). **(b)** Unit cell of Co-doped $EuFe_2As_2$, with ferromagnetic Eu layers spontaneously polarized out of plane. Stress is applied in the Fe–As bonding direction to tune tunes nematicity without detwinning the domains (figure 2(d)). The reorientation of Eu moments to align in-plane as an in-plane magnetic field is applied can be observed from X-ray magnetic circular dichroism (XMCD) results. The minimum resistivity is attained precisely at the field value where Eu moments are fully saturated in-plane, revealing the mechanism of field-induced superconductivity in this system. The applied stress strongly tunes the resistivity but has virtually no effect on the Eu moment saturation field or value (From [25]).

## 2.3 Magnetism probed by resonant magnetic scattering in the iridates

Magnetism can also be investigated by directly probing magnetic reflections via resonant X-ray magnetic scattering (RXMS). When combined with in situ anisotropic strain, this powerful technique can be used to continuously monitor the evolution of different magnetic structures while simultaneously tuning the scale of the induced anisotropic lattice distortion. This is well demonstrated in a study on the iridate compound $Sr_2IrO_4$.

Iridate is a representative *5d* electron system with a strong spin–orbit coupling (SOC), resulting in an exotic pseudospin-1/2 ground state [27] shown in figure 4(a). In condensed matter systems, electrons orbiting around the nucleus of an atom possess a degree of freedom of orbital momentum and spin. In certain cases, these two couplings result in a composite state. This so-called SOC can also be regarded as a relativistic interaction of the spin of a moving electron inside a potential. The strength of SOC is approximately proportional to $Z^4$, where Z is the atomic number. Hence, SOC is often of particular importance in 4d and 5d transition metal systems. One of the most well-known iridates is the quasi-2D $Sr_2IrO_4$, where the $IrO_2$ forms a layered structure separated by SrO [28, 29] (figure 4(b)). The combination of the Dzyaloshinskii–Moriya (DM) interaction and large SOC results in Ir pseudospin-canting, which in turn produces a net moment within each of the $IrO_2$ layers [30]. These canted moments order antiferromagnetically below ~230 K but can also be easily aligned by an external magnetic field. This metamagnetic transition is accompanied by a large magnetoresistance (MR) [31]. Hence, $Sr_2IrO_4$ an ideal prototype in the emerging field of AFM spintronics [32].

Furthermore, two additional key facts of the $Sr_2IrO_4$ system need to be discussed. First, although the DM interaction leads to an in-plane $IrO_6$ octahedra staggered rotation, its Hamiltonian sums to zero over a closed loop in a 2D picture [33]. In other words, the DM interaction does not make any remarkable contribution to the magnetic anisotropy in $Sr_2IrO_4$. Second, given that the $J_{eff} = 1/2$ orbit is not degenerated, one may consider a pseudospin-1/2 system to not host the Jahn–Teller (JT) effect. However, this consideration may not be suitable as the mixed $t_{2g}$ waveforms in the $J_{eff} = 1/2$ wave function can give rise to an effect similar to the classic JT effect [34]. This so-called "pseudo-JT effect" predicts a spontaneous structural orthorhombicity because of AFM ordering, to a scale of approximately $10^{-4}$ under a low energy limit [35]. To summarize, $Sr_2IrO_4$ is a quasi-2D system with strong SOC, and its weak anisotropies are critical in stabilizing its magnetic structure.

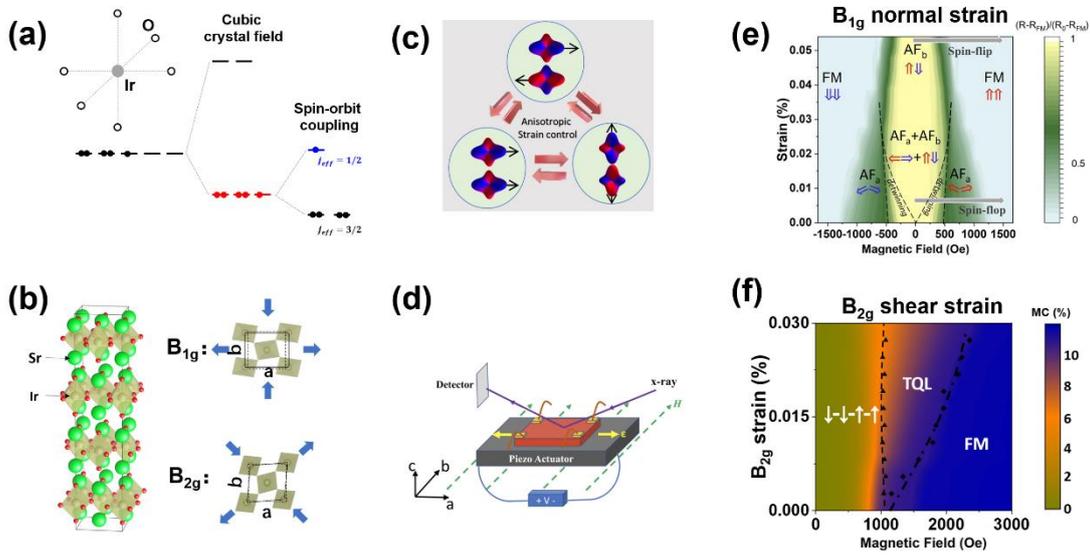

**Figure 4.** (**a**) Schematic showing the splitting of the *d*-orbit under a crystal field and strong SOC. (**b**) Schematic illustrating the control of the pseudospin waveform via anisotropic strain. (**c**) Crystal structure of $Sr_2IrO_4$. The effect of a $B_{1g}$ normal strain or a $B_{2g}$ shear strain on a single $IrO_2$ layer is shown on the right. (**d**) Schematic of the elasto-X-ray approach for studying metamagnetism under anisotropic strain. (**e**) Strain–magnetic field phase diagram of $Sr_2IrO_4$ at 210 K when a $B_{1g}$-type normal strain is applied in the ab plane. (**f**) Strain–magnetic field phase diagram of $Sr_2IrO_4$ at 210 K when a $B_{2g}$-type shear strain is applied in the *ab* plane.

Recent studies suggest that epitaxial growth has no notable effect on $Sr_2IrO_4$ [36]. However, the large SOC and intrinsic weak magnetic anisotropy afford a remarkable route to manipulate its magnetic/electric properties by applying an anisotropic strain (see the concept in figure 4(c)). Indeed, by applying an in situ anisotropic normal strain of <0.05%, the metamagnetic transition field of $Sr_2IrO_4$ can be modulated by approximately 300%, leading to a deterministic on- and off-switching of its magnetism under a fixed moderate magnetic field [7]. This result demonstrates an elasto-approach to accommodate the longstanding objective in the field of AFM spintronics for effective nonmagnetic control over the Neel order. To determine the origin, resonant X-ray scattering and transport with in situ anisotropic strain were simultaneously measured at beamline 4, Advanced Phonon Source (APS) of the Argonne National Laboratory (figure 4(d)). The measurements revealed that this drastic response originates from the complete strain tuning of the transition between the spin–flop and spin–flip limits and is always accompanied by large elastoconductance and magnetoconductance values (figure 4(e)).

In the work described above, the $B_{1g}$ anisotropy induced with such a small strain value has comparable energy levels with the $Sr_2IrO_4$ intrinsic $B_{1g}$ anisotropies, making the response of the AFM order to externally induced orthorhombicity remarkably efficient. Therefore, it is natural to consider introducing anisotropies of a different symmetry group. In a following work by Pandey et al. [22], a $B_{2g}$-type shear strain was applied to the $Sr_2IrO_4$ system where a thin piece of crystal was stretched in the [110] lattice direction. Elastoelectric study was combined with in situ RXMS, and the combined technique established the emergence of a new magnetic phase with broken translation symmetry (figure 4(f)). It is identified by further model analysis as a continuously tunable 12-layer spatial spin modulation along the c-axis. This behavior originates from the competition of anisotropic interactions of orthogonal irreducible representations, i.e., an unusual strain-activated anisotropic interaction of the fourth order that competes with the inherent quadratic anisotropic interaction. This behavior enables electrically controllable and electronically detectable metamagnetic switching despite the AFM insulating state. Meanwhile, the application of a 0.1% anisotropic strain increases the magnon energy of $Sr_2IrO_4$ by 40%, possibly opening up multifold design options for reconfigurable magnonic devices [37].

*2.4 Charge order probed by resonant scattering in copper-based superconductors and related materials*

A combination of tunable strain and X-ray techniques was successfully used to examine the charge order in copper-based high-temperature superconductors, often referred to as cuprate superconductors. The phase diagrams of many cuprates include a Neel AFM order that is suppressed by doping and the emergence of a charge density wave (CDW) order and high-temperature superconductivity with increased doping. Because the onset of superconductivity tends to suppress the CDW order, the two phases are generally regarded to compete with each other.

The key structural feature shared by many families of cuprates is the $CuO_2$ plane, in which Cu and O atoms form a square lattice. This plane hosts AFM order and charge order and is considered essential for superconductivity. While the iron-based superconductors and some of the simpler cuprates are structurally tetragonal, cuprates from the $ReBa_2Cu_3O_7$ (Re = Y, Nd, etc.) system are structurally orthorhombic because of the presence of CuO chains in the intermediary layers between the $CuO_2$ planes. In this system, a 2D biaxial CDW order forms; this order can be decomposed into two distinct order parameters aligned parallel or perpendicular to the CuO chain direction. The presence of a CDW can be determined by measuring the scattered X-ray intensity at a wave vector corresponding to the periodicity of the CDW order parameter. The strain tunability of this intensity can then yield important information about the electronic coupling to the lattice. Because the $ReBa_2Cu_3O_7$ crystal structure is already orthorhombic, strain is not a necessary tool for breaking the rotational symmetry or detwinning the structural/CDW domains. Instead, the application of uniaxial pressure can act to tune the orthorhombicity of the lattice and thus the CDW order.

The first study [23] that combined tunable uniaxial strain and inelastic scattering (IXS) was conducted on the underdoped compound $YBa_2Cu_3O_{6.67}$, which has a biaxial incommensurate in-plane 2D CDW order below T = 150 K. Uniaxial pressure was applied perpendicular to the CuO chain such that increasing the pressure (i.e., compressive strain) further enhanced orthorhombicity. X-ray scattering at the wavevector $q_{2d}$ = (0 0.31 6.5) (i.e., perpendicular to the stress direction) revealed the presence of the 2D CDW within the CDW + superconducting phase. The application of pressure caused a drastic increase in the elastic scattering intensity, demonstrating the strain enhancement of the CDW. Interestingly, the applied pressure also induced a uniaxial CDW with three-dimensional (3D) dispersion in a narrow temperature range centered near T = 50 K and measured at the wavevector $q_{3d}$ = (0 0.31 7). A comparison of the sharpness of the measured elastic peaks at $q_{2d}$ and $q_{3d}$ indicate a correlation volume that is more than two orders of magnitude larger for the strain-induced 3D order than for the 2D order. In other words, the applied pressure causes much larger sample volumes to form a coherent CDW state. Finally, from the inelastic scattering, the Kohn anomaly associated with the softening of a phonon mode was identified as the onset of the 3D CDW order. Thus, elastic and inelastic scattering together provide important information on the coupling of emergent CDW orders to the strain-tunable lattice.

Another study by the same group on $YBa_2Cu_3O_{6.67}$ [38] further advanced the experimental methodology; in this study, pressure was applied either in-plane perpendicular (*a* lattice constant) or parallel (*b* lattice constant) to the CuO chain direction. When incident X-rays tuned to the energy of the Cu L-edge were used, the RIXS measurement exhibited increased sensitivity to the Cu atoms in the $CuO_2$ planar layers. These modifications enabled a more careful examination of the unidirectionality versus bidirectionality of the 2D and 3D CDW orders. The key finding of this study was that pressure applied in the *a* (*b*) direction strongly enhanced the elastic intensity associated with the 2D CDW in the *b* (*a*) direction, with minimal impact on the intensity of the CDW in the *a* (*b*) direction. Thus, the *a*- and *b*-directional CDW orders have distinct CDW order parameters that evolve independently under applied pressure. This result rules out a "checkerboard" CDW order in favor of distinct uniaxial CDW orders that coexist in the *a* and *b* directions. However, while the pressure applied along *a* can induce a 3D CDW along *b*, the pressure along *b* did not generate a 3D CDW along *a*. These combined strain and X-ray results thus demonstrate a genuine rotational inequivalence in the structural–electronic coupling.

Yet another study [39] explored the strain tunability of the CDW order in $ErTe_3$, a material considered to be a toy model system of cuprates owing to its similar crystal structure and simpler phase diagram. $ErTe_3$ is fundamentally orthorhombic because of the presence of an ErTe glide plane; however, the longer in-plane *c* lattice constant is only 0.1% larger than the shorter in-plane *a* lattice constant. A unidirectional CDW order spontaneously forms below 270 K in the longer *c* direction. In this study, tension was applied in the *a* direction to flip the orthorhombicity of the lattice, i.e., to make *a* longer than *c*. In doing so, the CDW intensity along *c* reduced monotonically with increasing tension along *a*, whereas a new CDW peak along *a* emerged under sufficiently large tension. Thus, by applying the appropriate strain, the direction of the CDW could be switched. Furthermore, at an intermediate tension value where $a \approx c$, the system demonstrated an emergent electronic tetragonality despite its structural orthorhombicity. Although previous studies [40] on elastoresistance and elastocaloric measurements had indicated the switchability of the CDW direction and this emergent tetragonality, by a combination of strain and XRD techniques, these features were definitively demonstrated.

## 3. Outlook

The examples presented above underscore the efficacy of integrating in situ anisotropic strain with synchrotron X-ray techniques. This strategic combination not only resolves elasto-structural modulations that are crucial for a comprehensive understanding of emergent phenomena but also enables a direct correlation between the macroscopic behaviours, such as magnetoresistance, nematicity and superconductivity, to the microscopic mechanism like lattice distortion and spin ordering. This unique capability not only elucidates the intricate interplay among different order parameters but also disentangles their often-complex relationships. Looking ahead, the elasto-X-ray approach is poised to become indispensable in quantum materials research, and have profound and enduring impacts. We anticipate widespread applications of the elasto-X-ray techniques in various representative systems, further solidifying its role as a transformative tool in the exploration of quantum materials.

The first kind is geometric frustration which occurs in certain types of crystal lattices where the energy minimization cannot be simultaneously satisfied at all lattice points [41]. The resulted large degree of degeneracy may trigger many interesting phenomena, from observations of topological magnetic structure [42] to quantum entangled spin liquid states [43]. A conventional way to manipulate frustrated spins is by applying an external magnetic field and/or high pressure, which are however not suitable for device applications. The use of a tunable anisotropic strain could add a new dimension by inducing a different symmetry-breaking control, while X-ray scattering can be employed to probe the change in both the frustrated crystalline and magnetic lattices. Indeed, recent studies on the $Mn_3Sn$ Kagome system have already revealed a large piezomagnetic effect that switches the sign of the Hall signal by an anisotropic strain less than 0.1% [44]. The technique could be readily extended to other frustrated systems such as centrosymmetric skyrmion [45], (artificial) spin ice [46], and quantum spin liquid candidate [47]. For example, in the skyrmion texture that is a superposition of different spin modulation patterns, the relative weight of the modulations at different wave vector *q* could in principle be tuned by anisotropic strain that lifts their degeneracy. Such effect can be directly verified by *in situ* resonant X-ray magnetic scattering. Furthermore, for candidate materials of quantum spin liquid/ice, the potential integration of elasto-X-ray and elasto-nuclear magnetic resonance (NMR) could even afford a revolutionary approach by continuous tuning the lattice distortion and trace any long-range magnetic ordering at the same time.

Another major class of systems of interest is the atomically thin 2D materials. In this type of systems, samples can withstand much larger strain than bulk crystals. Elastic strain is therefore a highly effective approach to manipulate the physical properties. For instance, membrane samples of $SrTiO_3$ and $La_{0.7}Ca_{0.3}MnO_3$ have already show remarkable responses to anisotropic strain [48,49]. A strain-induced antiferromagnetic to ferromagnetic transition has been realized in atomically thin CrSBr [50]. Elasto-X-ray techniques with the bright synchrotron beams could provide critical insights into these systems despite the small sample volume. In the meantime, there are also studies utilizing in situ anisotropic strain with other measurements, such as Raman spectroscopy, specific heat, NMR [51] and so on. Some of these techniques could be compatible with synchrotron X-ray and carried out simultaneously with elasto-X-ray measurements. Sample environment with such multi-modal capabilities would bring research of elasto-quantum materials to a new era.

The elasto-X-ray approach can also be beneficial in the quickly growing area of altermagnetism, where the combination of antiferromagnetism and alternating structural motifs form a unique spin-lattice pattern [52]. The coupling between the spin alternation and the crystal symmetry makes this type of magnetic materials an idea playground for lattice-driven emergent phenomena. Specifically, strain/stress can induce and control ferromagnetic behaviours of the antiferromagnetic order, such as Hall effects, by providing the necessary parity-time-reversal-symmetric perturbation [53]. Direct experimental proof of altermagnetism thus requires probing both the crystal and magnetic structures in conjunction with transport measurements under in situ strain tuning [54]. Therefore, the elasto-X-ray technique holds the potential to provide an efficient means of manipulating the altermagnetic properties, simultaneously allowing for the direct monitoring of distinctive elasto-responses at the atomic level.

## 4. New Opportunities for Advanced Photon Sources

There are in fact great upsides for elasto-X-ray techniques simply due to the development of the new generation synchrotron source, such as the APS upgrade (APS-U). For instance, at the new POLAR beamline (4-ID) of the APS-U, an integrated strain platform will be available for hard X-ray spectroscopy and diffraction measurements under magnetic field with a highly coherent and highly focused (100nm for resonant diffraction and low-field (B=2T) spectroscopy, 400nm for high-field (B=9T) spectroscopy). Such small spot sizes will enable new precision strain tuning of the lattice and lattice-coupled phenomena nanometric spatial resolution. For instance, previous detwinning studies [1, 4] used beams of spot size >50 μm, which results in a scattered X-ray intensity that includes contributions from many structural domains. A small beam size will enable the tracking of individual structural and magnetic domains as a function of strain. Real-space imaging of the strain-tunable domain structure will be enabled via raster scanning of the small beam spot over the sample surface and by new ptychography techniques. The highly coherent beam will also enable studies of few-layer-thin flake samples and ultrathin films, which have rarely been investigated using X-ray techniques [55]. These new possibilities will ensure that the range of research questions that can be addressed using elasto-X-ray techniques will only grow with time.

## Acknowledgements

J. L. acknowledges support from the National Science Foundation under Grant No. DMR-1848269 and the Air Force Office of Scientific Research under Grant No. FA9550-23-1-0502. H. Z. acknowledges support from the National Science Foundation of China (Grant No. 12204068), the Natural Science Foundation of Jiangsu Province (Grant No. BK20220616), and the Jiangsu Higher Education Research Program of Base Science under Grant No. 22KJB140001.